\documentclass[12pt]{article}

\def\la{\lambda}

\def\om{\omega}

\def\frac#1#2{{\textstyle{{#1}\over {#2}}}}

\def\vev#1{\langle {#1}\rangle}

\def\lsim{\mathrel{\rlap{\lower4pt\hbox{\hskip1pt$\sim$}}
   \raise1pt\hbox{$<$}}}
\def\gsim{\mathrel{\rlap{\lower4pt\hbox{\hskip1pt$\sim$}}
   \raise1pt\hbox{$>$}}}
\def\sqr#1#2{{\vcenter{\vbox{\hrule height.#2pt
        \hbox{\vrule width.#2pt height#1pt \kern#1pt
        \vrule width.#2pt}
        \hrule height.#2pt}}}}

\def\pt#1{\phantom{#1}}

\def\nsc#1#2#3{\om_{#1}^{{\pt{#1}}#2#3}}

\newcommand{\beq}{\begin{equation}}
\newcommand{\eeq}{\end{equation}}
\newcommand{\bea}{\begin{eqnarray}}
\newcommand{\eea}{\end{eqnarray}}
\newcommand{\bit}{\begin{itemize}}
\newcommand{\eit}{\end{itemize}}

\begin{document}

\begin{center}
{\bf CONSEQUENCES OF SPONTANEOUS LORENTZ \\
VIOLATION IN GRAVITY}\footnote{Presented at the 
13th Marcel Grossmann Meeting,
Stockholm, Sweden, July 2012}
\end{center}

\begin{center}
{Robert Bluhm \\ }
{Physics Department \\ }
{Colby College \\ }
{Waterville, ME 04901 USA\\}
\end{center}

\begin{abstract}
\noindent
A brief summary of some of the main consequences
of spontaneous Lorentz violation in gravity is presented,
including evasion of a no-go theorem,  
concomitant spontaneous diffeomorphism breaking,
the appearance of massless Nambu-Goldstone modes
and massive Higgs modes,
and the possibility of a Higgs mechanism in gravity.
\end{abstract}

\section{Introduction}	

There has been a great deal of interest in the possibility of Lorentz violation
in recent years stemming from ideas in quantum gravity, modified gravity theories,
massive gravity, and cosmological models attempting to explain dark energy
and/or dark matter.
There have also been significant improvements
in experimental tests of Lorentz violation,
and sensitivities at levels involving suppression 
by the Planck scale have been attained.
The theoretical framework known as the
Standard-Model Extension\cite{sme} 
(SME) consists of the
most general observer-independent effective field theory
incorporating Lorentz violation,
and it is routinely used by both theorists and experimentalists
to study and obtain bounds on possible 
forms of Lorentz violation.\cite{smereview,data}

While the SME can accommodate both explicit forms of Lorentz
breaking at the level of effective field theory as well as the 
process of spontaneous Lorentz violation,
there are differences that arise between these types
of Lorentz breaking in the context of gravity.
This summary looks at distinguishing these differences and
in particular what some of the consequences are
of spontaneous Lorentz violation in the context of gravity.

\section{Local Lorentz Symmetry and Diffeomorphisms}	

In the presence of gravity,
Lorentz symmetry is a local symmetry that holds
in local inertial frames at every spacetime point.
For this reason,
it is useful to use a vierbein formalism to investigate
Lorentz violation in the context of gravity.
The vierbein, $e_\mu^{\phantom{\mu}a}$,
connects tensor components in local Lorentz frames
(labeled using Latin indices) with tensor components in
the space-time frame (labeled using Greek indices).
The SME can be written using a vierbein formalism.

The SME Lagrangian consists of the most general
scalar function under local Lorentz and diffeomorphism
transformations that can
be formed using gravitational fields, particle fields
in the Standard Model, and 
Lorentz-violating coefficients known as SME coefficients.
In SME models with explicit Lorentz breaking,
the SME coefficients can be viewed as fixed background fields.
However, in models arising from a process of 
spontaneous Lorentz violation,
the SME coefficients are understood to be vacuum
expectation values (vevs) of dynamical tensor fields.
In either case, SME models with these forms of
Lorentz breaking can be investigated experimentally,
and bounds on the SME coefficients can be obtained.
However, there are effects that arise that
can further distinguish these types of symmetry breaking.

If the Lorentz violation is due to spontaneous breaking,
then nonzero tensor-valued vacuum values,
e.g., $\vev{T_{abc\cdots}}$, occur in the local Lorentz frames.
However, in a theory with spontaneous Lorentz breaking,
the vierbein also has a vacuum value, 
$\vev{e_\mu^{\phantom{\mu}a}}$.
When appropriate products of the vierbein vev act on
the local tensor vevs, 
the result is that tensor vevs,
e.g., $\vev{T_{\la\mu\nu\cdots}}$, also appear in
the spacetime frame.
These tensor vevs in the spacetime frame spontaneously
break local diffeomorphisms.
Conversely, if a vev in the spacetime frame spontaneously
breaks diffeomorphisms,
then the inverse vierbein acting on it gives rise to vevs
in the local frames
that spontaneously break local Lorentz symmetry.
Hence spontaneous local Lorentz
breaking implies spontaneous diffeomorphism breaking
and vice versa.\cite{rbak}

Note that this concomitant symmetry breaking 
does not occur for the case of explicit symmetry breaking.
For example, a Fierz-Pauli model describing 
massive metric excitations $h_{\mu\nu}$
in a Minkowski background has explicit
diffeomorphism breaking,
but the theory does not break Lorentz symmetry.
Similarly, one can write down models that explicitly break
local Lorentz symmetry, e.g., using products of the spin connection,
$\nsc \mu a b$, which are not Lorentz tensors,
while maintaining diffeomorphism invariance. 

In the gravity sector of the SME,
a no-go theorem shows that when explicit Lorentz breaking occurs,
an inconsistency can arise between conditions stemming from
the field variations and symmetry considerations 
with geometrical constraints that must hold, 
such as the Bianchi identities.\cite{ak}
However, it was also shown that the case of
spontaneous Lorentz breaking evades the no-go theorem.
The difference is that in a theory with explicit breaking,
the SME coefficients are not associated with dynamical fields;
while in the case of spontaneous Lorentz breaking they are.
This creates a difference in the conditions that must hold with explicit
symmetry breaking compared to spontaneous symmetry breaking.
A consequence of the no-go theorem is that the gravity sector
of the SME can only avoid incompatibility 
with conventional geometrical constraints if the symmetry breaking is spontaneous.

\section{Spontaneous Lorentz Violation}

Spontaneous breaking of Lorentz and diffeomorphism symmetry implies that 
massless Nambu-Goldstone (NG)
modes should appear (in the absence of a Higgs mechanism).
In general, there can be up to as many NG modes as there are broken symmetries. 
Since the maximal symmetry-breaking case would yield six broken Lorentz generators 
and four broken diffeomorphisms,
there can be up to ten NG modes.
A natural gauge choice is to put all the NG modes into the vierbein,
and a simple counting argument shows that this is possible.
With no spontaneous Lorentz violation, 
the six Lorentz and four diffeomorphism degrees
of freedom can be used to reduce the vierbein from 16
down to six independent degrees of freedom.  
However, in a theory with spontaneous Lorentz breaking,
alternative gauge-fixing choices can be made so that
the NG modes can naturally
be incorporated in the vierbein.
Of course, 
some of the NG modes might appear as ghosts,
and it is for this reason that most models with spontaneous Lorentz breaking
involve vevs that break fewer than ten spacetime symmetries.

In theories of spontaneous Lorentz breaking, 
the symmetry breaking is usually induced by a potential term in
the Lagrangian that has a degenerate minimum space.
The NG modes are excitations away from the vacuum
that stay in the minimum space,
while massive Higgs modes are excitations that go
up the potential well away from the minimum.  
In conventional gauge theory,
the fields in the potential are scalar fields,
and the Higgs modes do not involve the gauge fields.
However, with spontaneous Lorentz breaking,
the metric typically appears in the potential,
and for this reason massive Higgs modes can occur
for the metric excitations in a process that has no
analog in the case of conventional gauge theory.

In gravitational theories,
since the symmetry breaking is local,
the possibility of a Higgs mechanism occurs as well.
In a Higgs mechanism,
the would-be NG modes become additional degrees 
of freedom for massive gauge fields.
In theories with spontaneous Lorentz breaking,
since diffeomorphisms are also spontaneously broken,
there are two potential Higgs mechanisms.
The gauge fields associated with diffeomorphisms 
and Lorentz symmetry are the metric excitations and the spin connection.
However,
a Higgs mechanism involving the metric does not occur.\cite{ks}  
This is because the mass term that is generated by covariant derivatives
involves the connection,  which consists of derivatives of the metric
and not the metric itself.
However, for the case of Lorentz symmetry, 
a conventional Higgs mechanism can occur.\cite{rbak}
Here, the relevant gauge fields (for the Lorentz symmetry) 
are the spin connection.  
These appear directly in covariant derivatives acting 
on local tensor components,
and when the local tensors acquire a vev,
quadratic mass terms for the spin connection can be generated.
Note, however, a viable Higgs mechanism involving the spin connection can 
only occur if the spin connection is a dynamical field.  
This then requires that there is nonzero torsion and 
that the geometry is Riemann-Cartan.

\vfill


\begin{thebibliography}{00}

\bibitem{sme}
V.A.\ Kosteleck\'y and R.\ Potting,
Phys.\ Rev.\ D {\bf 51}, 3923 (1995);
D.\ Colladay and V.A.\ Kosteleck\'y,
Phys.\ Rev.\ D {\bf 55}, 6760 (1997);
Phys.\ Rev.\ D {\bf 58}, 116002 (1998);

\bibitem{smereview}
R.\ Bluhm, arXiv:1302.1150;

\bibitem{data}
V.A.\ Kosteleck\'y, and N.\ Russell,
Rev.\ Mod.\ Phys.\ {\bf 83}, 11 (2011);
arXiv:0801.0287.

\bibitem{rbak}
R.\ Bluhm and V.A.\ Kosteleck\'y,
Phys.\ Rev.\ D {\bf 71}, 065008 (2005);
R.\ Bluhm, S.-H.\ Fung, and V.A.\ Kosteleck\'y,
Phys.\ Rev.\ D {\bf 77}, 065020 (2008).

\bibitem{ak}
V.A.\ Kosteleck\'y,
Phys.\ Rev.\ D {\bf 69}, 105009 (2004).

\bibitem{ks}
V.A.\ Kosteleck\'y and S.\ Samuel,
Phys.\ Rev.\ D {\bf 40}, 1886 (1989).




\end{thebibliography}
\end{document}